\documentclass[aip,apl,reprint,floatfix]{revtex4-1}
\usepackage{eqnarray,amsmath}
\usepackage{graphicx}
\usepackage{dcolumn}
\usepackage{bm}
\usepackage{wasysym}
\usepackage{gensymb}
\usepackage{color}
\definecolor{amber}{rgb}{1.0, 0.49, 0.0}
\usepackage[normalem]{ulem}
\usepackage{calrsfs}
\usepackage{float}

\DeclareMathAlphabet{\pazocal}{OMS}{zplm}{m}{n}

\definecolor{light-gray}{gray}{0.45}
\definecolor{deeogreen}{RGB}{34,159,34}
\begin{document}
\title{Eddy-current imaging with nitrogen-vacancy centers in diamond}
\author{Georgios Chatzidrosos}
\email{gechatzi@uni-mainz.de}
\affiliation{Johannes Gutenberg-Universit{\"a}t  Mainz, 55128 Mainz, Germany}
\author{Arne Wickenbrock}
\affiliation{Johannes Gutenberg-Universit{\"a}t  Mainz, 55128 Mainz, Germany}
\author{Lykourgos Bougas}
\affiliation{Johannes Gutenberg-Universit{\"a}t  Mainz, 55128 Mainz, Germany}
\author{Huijie Zheng}
\affiliation{Johannes Gutenberg-Universit{\"a}t  Mainz, 55128 Mainz, Germany}
\author{Oleg Tretiak}
\affiliation{Johannes Gutenberg-Universit{\"a}t  Mainz, 55128 Mainz, Germany}
\author{Yu Yang}
\affiliation{University of Science and Technology of China (USTC), Hefei 230026, China}
\author{Dmitry Budker}
\affiliation{Johannes Gutenberg-Universit{\"a}t  Mainz, 55128 Mainz, Germany}
\affiliation{Helmholtz Institut Mainz, 55099 Mainz, Germany}
\affiliation{Department of Physics, University of California, Berkeley, CA 94720-7300, USA}
\affiliation{Nuclear Science Division, Lawrence Berkeley National Laboratory, Berkeley, CA 94720, USA}
\date{\today}

\begin{abstract}
We demonstrate microwave-free eddy-current imaging using nitrogen-vacancy centers in diamond. By detecting the eddy-current induced magnetic field of conductive samples, we can distinguish between different materials and shapes and identify structural
defects. Our technique allows for the discrimination of different materials according to their conductivity. The sensitivity of the measurements is calculated as 8$\times 10 ^{5}$\,S/m\,$\sqrt[]{\textrm{Hz}}$ at 3.5\,MHz, for a cylindrical sample with radius $r_0$\,=\,1\,mm and height $h$\,=\,0.1\,mm (volume $\sim$\,0.3\,mm$^3$), at a distance of 0.5\,mm. In comparison with existing technologies, the diamond-based device exhibits a superior bandwidth and spatial resolution. In particular, we demonstrate a flat frequency response from DC to 3.5 MHz and a spatial resolution of 348\,$\pm$\,2\,$\mu$m.
\end{abstract}

\maketitle
\section*{Introduction}
Magnetic induction measurements have proven to be useful in biomedicine\cite{Marmugi2016}, security and surveillance\cite{Deans2018,Hussain2015}, and materials testing\cite{Angelo2015}.
%and their compatibility with modern technologies, like machine learning\,\cite{DeansMachine2018}
Magnetic-induction imaging works by detecting magnetic fields produced by eddy-currents. When a material is placed in an alternating magnetic field (primary field), electric fields will be induced in the material, causing eddy-currents to flow. These in turn will produce a secondary magnetic field, that can be detected along with the primary field.
The secondary field depends on the material's properties and shape, as well as the skin depth of the primary field. Eddy-current detection has been commercially available with coil-based devices (e.g. Zetec, MIZ-22). Recently it has also been demonstrated using vapor-cell magnetometers\cite{Arne2016}.

%Since their first demonstrations\cite{Korjenevsky2000,Al-Zeibak1995}, most of the application involving these techniques have utilized atomic vapor-cell magnetometers\cite{Deans2018, Hussain2015,Marmugi2016, DeansMachine2018,Arne2016}. 

Here, we demonstrate eddy-current imaging using magnetic sensing with nitrogen-vacancy (NV) centers in diamond. Compared to other sensors, diamond-based devices operate in a wide temperature range and may have nanoscale-resolution\cite{Balasubramanian2008,Maze2008,Rittweger2009} with high sensitivity\cite{Barry2016,Chatzidrosos} and wide bandwidth\cite{Chatzidrosos}. %NV magnetometers have been used to perform single neuron-action potential detection\,\cite{Barry2016}, single protein spectroscopy\,\cite{Lovchinsky2016}, as well as in vivo thermometry \cite{SingleNVThermometry}.  
%In the case of eddy-current detection, this allows identification of various kinds and thicknesses of materials. The small size of NV sensors allows for close proximity which, combined with their nanoscale resolution, can be used to identify small/localized defects in the imaged samples.
The present eddy-current measurements are performed using an all-optical NV magnetometer. The device has a bandwidth of 3.5\,MHz and exhibits a spatial resolution of 348\,$\mu m$.

\section*{Experimental setup}
\begin{figure}[htp]
\centering
\includegraphics[width=\linewidth]{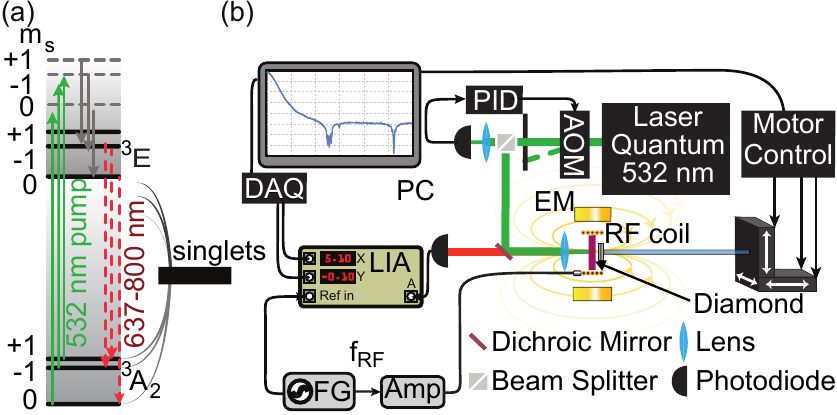}
\caption{(a) Relevant NV center energy levels and transitions. Solid green and red lines indicate excitations, dashed lines indicate radiative transitions, and gray solid lines indicate non-radiative transitions.
(b) Schematic of the experimental setup.}
\label{fig:fig1}
\end{figure}
In Fig.\,\ref{fig:fig1}\,(a) we present the NV center energy levels. The ground and excited electronic spin-triplet states of the NV are $^{3}$A$_{2}$ and $^{3}$E, respectively, with the transition between them having a zero-phonon line at 637\,nm. 
% Present the energy levels. singlet.
Additionally, there are two singlet states.
% Explaining the picture: lines. Spin dependence.
Optical transition rates in the system are spin-independent, however, the probability of nonradiative intersystem crossing from $^3$E to the singlets is several times higher for $m_s=\pm 1$ than that for $m_s=0$\cite{Dumeige2013}. 
% Optical pumping. Step one Prepare
As a consequence, under continuous illumination with green pump light (532\,nm), NV centers are prepared in the $^{3}$A$_{2}$ $m_s=0$ ground state sublevel and in the metastable $^{1}$E singlet state. Population in the $^{3}$A$_{2}$ $m_s=0$ state, then gets excited by the green light and decays back, emitting red photoluminescence (PL).

The setup consists of a microwave-free diamond magnetometer in AC-mode (discussed in detail in the following paragraph), a driving coil to induce eddy currents and a 3D-translation stage to make spatially dependent eddy-current measurements i.e. to create a conductivity image.
The setup is shown in Fig.\,\ref{fig:fig1}\,(b). The diamond  used for this sensor is a type Ib, (111)-cut, HPHT grown sample, purchased from Element Six. Its dimensions are (3$\times$3$\times$0.400\,mm$^3$. The initial nitrogen concentration of the sample was specified as $<$\,110\,ppm. The sample was irradiated with 14\,MeV electrons at a dosage of $10^{18}\,\textrm{cm}^{-2}$ and then annealed at 700\,$^o$C for three hours.
% lasers we use green light.
Green light is provided to the setup by a diode-pumped solid-state laser (Laser Quantum, gem 532).
% What we do to Green light
The power of the laser is stabilized using an acousto-optical modulator (AOM, ISOMET-1260C with an ISOMET 630C-350 driver) controlled with a proportional-integral-derivative controller (PID, SRS SIM960). The light is focused on the diamond using a 50\,mm lens.
% Fluorescence collection
Before being detected with a photodiode (Thorlabs PDA36A-EC), the PL is filtered with a dichroic mirror.
%Translation stage
The samples to image are attached on to a 24\,cm non-conductive rod, which is screwed onto a motorized, computer controlled, 3D-translation stage system (Thorlabs, MTS25-Z8). Using the translation stage, the samples are moved in front of the diamond.
% Coils and Magnet
The eddy-current inducing magnetic field is produced with a function generator (FG, Tektronix AFG 2021), ranging from 2\,Hz to 4\,MHz. The signal is supplied to a coil placed around the diamond (RF coil) with five turns and 3.4\,mm diameter. It is made from 0.05\,mm diameter copper wire.
% Magnet
Both the diamond and the coil are placed inside the bore of a custom-made electromagnet (EM). The magnet consists of $\sim$200\,turns wound with a rectangular-cross-section (1.4\,mm$\times$0.8\,mm) wire around a 5\,cm-diameter bore. The coil is wound on a water-cooled copper mount, and produces a background field of 2.9\,mT per ampere supplied. A current up to 55\,A is provided by a Keysight N8737A power supply.
A lock-in amplifier (LIA) (SRS, SR865), referenced at the eddy-current frequency, detects the amplitude ($R$) and phase ($\theta$) of the PL intensity modulation. $R$ and $\theta$ are recorded on a computer (PC) along with the position of the 3D-translation stages.

\section*{Magnetometry}

\begin{figure}[htp]
\centering
\includegraphics[width=\linewidth]{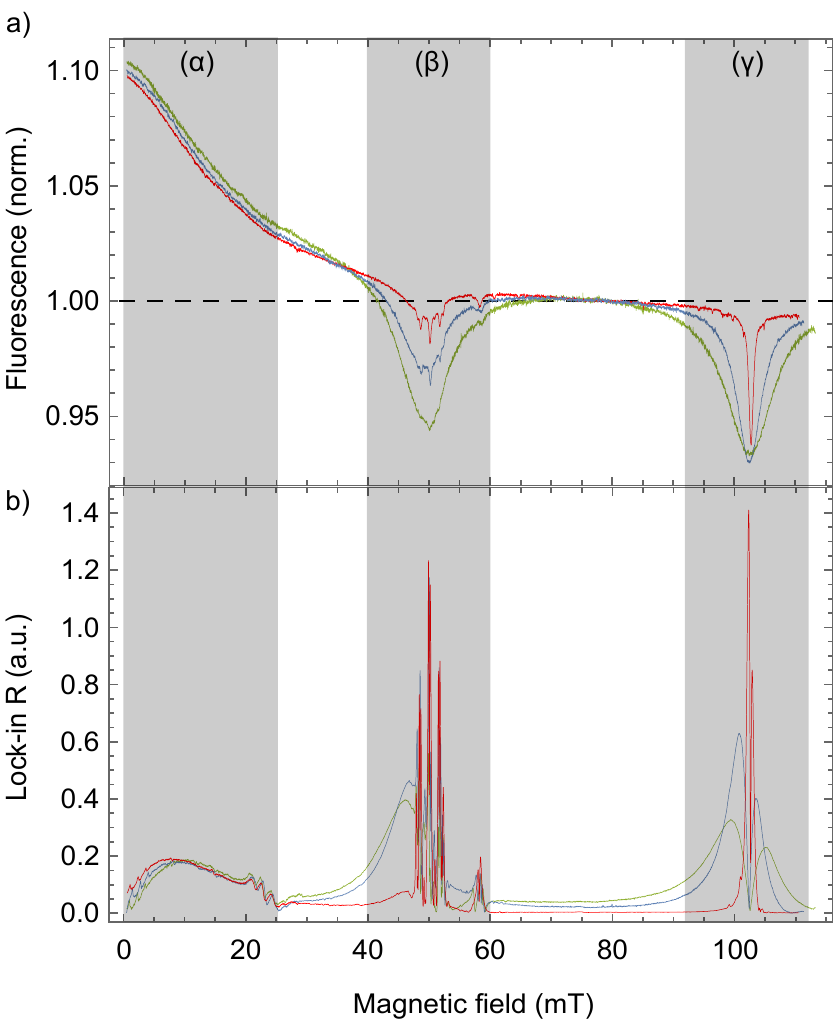}
\caption{(a) Fluorescence as a function of magnetic field at different alignments (up to $\sim$\,5$^o$) between the NV and magnetic field axis. The smallest misalignment is represented by the red trace and the biggest by the green. 
The data is normalized to the PL at 80\,mT. (b) Lock-in amplifier $R$ component for the same alignments between the NV and the magnetic field axis, with an applied magnetic field modulation of 60\,kHz and an amplitude of 50 $\mu$T. The shaded areas $\alpha,\beta,\gamma$ represent magnetic field values at which we performed AC magnetometry}
\label{fig:fig2}
\end{figure}

The magneto-metric scheme we followed for this measurements enables all-optical detection of AC fields. Figure\,\ref{fig:fig2}\,(a) presents normalized PL measurements as a function of the background magnetic field at different alignments of the magnetic field relative to the NV axis. Figure\,\ref{fig:fig2}\,(b) shows the corresponding amplitude of the PL oscillation due to the oscillating magnetic fields, $R$ as measured with a LIA component from the lock-in amplifier. The traces display several features discussed in the literature\cite{Armstrong1,Hall2016,wickenbrock2016microwave, Ivanov2016, Wood2016_2,Zheng2017}. We investigated three different areas: $\alpha$) the slope from 0 to $\sim$\,25\,mT, $\beta$) the cross-relaxation features around 50\,mT and $\gamma$) the ground-state level anti-crossing (GSLAC) feature at 102.4\,mT. To perform the measurements, we apply a bias field and maximize $R$ for the different features. 
%While the signal is maximized we are insensitive to DC fields and sensitive to AC components.
While $R$ is maximized due to the quadratic shape of the feature, we are not sensitive to small changes in the bias field. However because we are demodulating at the frequency of the eddy current using a lock-in amplifier, we are sensitive to AC field changes.

As it is shown in Fig.\,\ref{fig:fig2}\,(b) the maximum $R$ is obtained at the GSLAC for the red trace. However, as demonstrated in the literature\cite{Zheng2017}, this feature is sensitive to misalignment. To make the sensor more robust to misalignment, we perform the measurements, at area $\alpha$. Even though area $\alpha$ is less sensitive it allows for robustness which will be useful in a portable device. Instead of requiring a highly homogeneous magnetic field, a standard permanent magnet could bias the field. The smaller field value additionally facilitates implementation in a miniaturized sensor and requires less power dissipation if an EM is used.

\begin{figure}[htp]
\centering
\includegraphics[width=8.6cm]{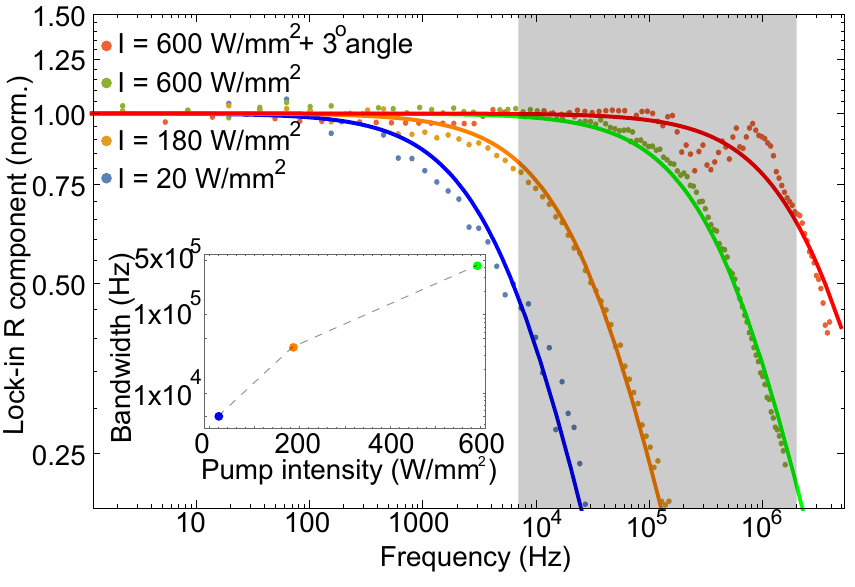}
\caption{Eddy current response as a function of frequency. The blue, orange and green measurements show the bandwidth of the device for different pump power intensities. The red measurement is taken at saturation pump intensity and with 3$^o$ misalignment between the NV and magnetic field axis. The shaded area between 7\,kHz and 2\,MHz represents the maximum reported bandwidth of current atomic based sensors used for eddy-current imaging.\cite{Arne2016,Marmugi22018}. The inset shows the bandwidth of the magnetometer versus intensity with the corresponding colors of the traces.}
\label{fig:fig3}
\end{figure}

For eddy-current detection measurements a broad bandwidth is important because the frequency of the primary field determines its penetration depth into the sample under study. The penetration depth in turn provides information about the geometry, thickness, and material of the sample. %Matching the skin depth with the thickness of the material, the field response can be optimized. 
Hence, to optimize images of different thickness materials, a wide frequency range is beneficial. Furthermore, higher frequencies enhance the response and make it possible to image materials with lower conductivity.
%When the penetrating depth matches the thickness of the material the signal is maximized\cite{Arne2016}. 
%With high bandwidth we are able to image thinner and less conductive materials.  
Figure\,\ref{fig:fig3}\ shows the normalized eddy-current response as a function of frequency, to characterize the sensor's bandwidth. The blue trace is taken at a pump intensity of I\,=\,20\,W/mm$^2$, the orange at I\,=\,180\,W/mm$^2$ and the green at I\,=\,600\,W/mm$^2$, which corresponds to the saturation intensity for the probed NV volume. The inset shows the normalized PL of the NV as a function of pump light intensity. The intensity values at which the data are taken are represented in the inset with the corresponding colors of the main plot. The measurements are fitted with first order low-pass-filter functions. As illustrated in Fig.\,\ref{fig:fig3}, the bandwidth can be tuned by varying the laser-light intensity and by changing the alignment between the NV axis and magnetic field, which affects the mixing of the NV m$_s$\,=\,0 and m$_s$\,=\,-1 states\cite{Zheng2017}. We demonstrate a maximum bandwidth $\sim$\,3.5\,MHz.

\section*{Spatial Resolution}

\begin{figure*}[htp]
\centering
\includegraphics[width=\textwidth]{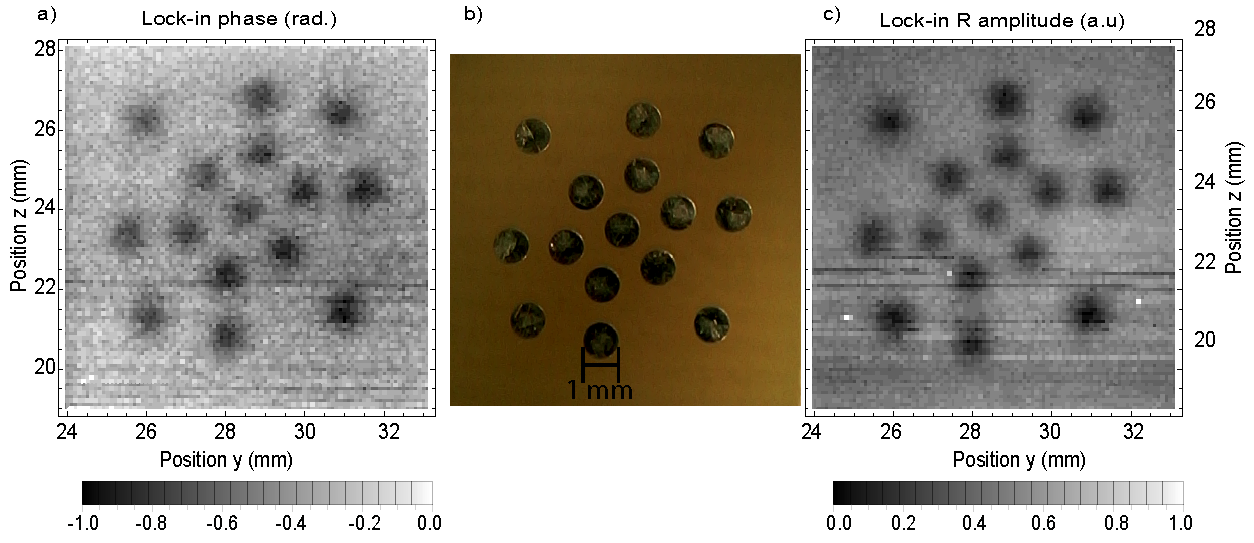}
\caption{a),c) Lock-in Phase and Lock-in $R$ amplitude for a PCB containing fifteen aluminum dots imaged with eddy-current b) photograph of the PCB with the corresponding length of 1\,mm noted for scale.}
\label{fig:Sfig2}
\end{figure*}

\begin{figure}[htp]
\centering
\includegraphics[width=8.6cm]{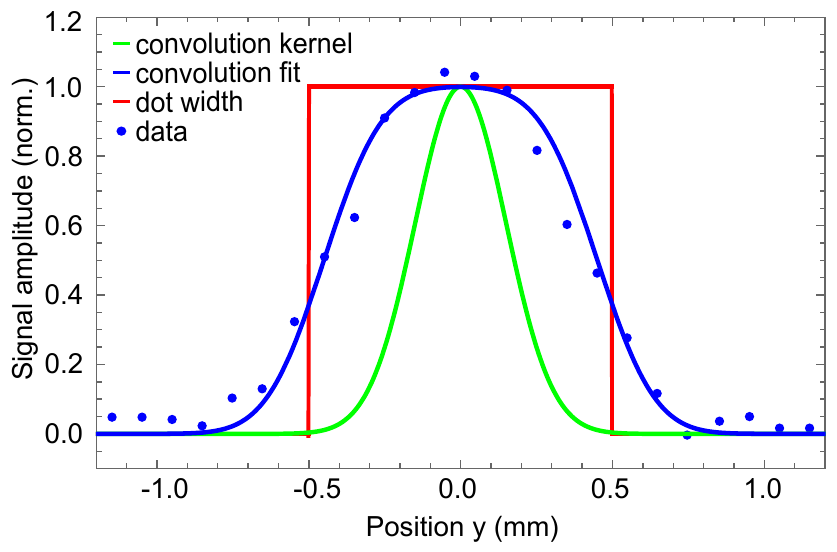}
\caption{Determining spatial resolution. The blue dots signify the average cross section of fifteen 1\,mm dots imaged by our sensor, the red plot signifies the dots' width as a square function, the green trace represents a Gaussian function needed as a kernel to recreate the experimental data and blue trace represents the convolution result of the red and green trace.}
\label{fig:fig4}
\end{figure}

One of the advantages of NV-based sensors is their spatial resolution. For eddy-current imaging, for a constant conductivity, the smaller a material is, the smaller the amplitude of the secondary field it produces. As a test of spatial resolution, we image fifteen 1\,mm-diameter 35\,$\mu$m thick dots made out of aluminum. The amplitude and phase response of the magnetic field sensor caused by these dots is shown in Fig.\,\ref{fig:Sfig2}. Both the dots and the distance between them, which is sub-mm, are clearly visible. Note that covering the pattern with aluminum foil adds an offset but does not significantly alter the eddy-current image patterns shown in Fig.\,\ref{fig:Sfig2} a and c.%

Figure\,\ref{fig:fig4}\ shows the average cross section of the fifteen 1 mm dots imaged in Fig.4. Using the average detected cross section data we deconvolute them with a square function, searching for a Gaussian function kernel that would recreate the experimental result. The kernel provides information about the
dots' width. In Fig.\,\ref{fig:fig4}\, the red plot signifies the dots' width of 1\,mm as a square function, the blue dots are the experimental data, the green trace represents the Gaussian function kernel and the convolution result of the red and green traces. All the traces amplitudes are normalized to unity for representation in the figure. 
The Gaussian kernel full-width-half-maximum is 348\,$\pm$\,2\,$\mu$m which we use as a measure for the spatial resolution of the sensor. The spatial resolution is mostly limited by the distance between sample and sensor. The diamond thickness sets the minimum distance to the sample. To decrease the distance between the sensor and the imaged sample a thinner diamond sample or a diamond with a shallow implanted NV layer could be used. A close proximity would improve measurement contrast and therefore the spatial resolution and allow the imaging of smaller objects.

\begin{figure*}
\centering
\includegraphics[width=\textwidth]{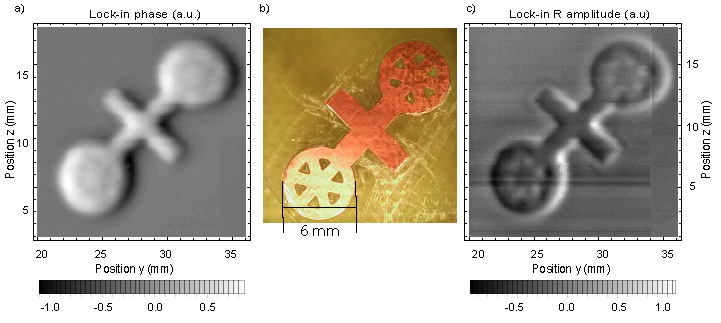}
\caption{\small{a),c) Lock-in $\theta$ and $R$ of a PCB depicting the symbol of the city of Mainz imaged with eddy-current. b) photograph picture of the PCB with the corresponding length of 6\,mm noted for scale.}}
\label{fig:fig5}
\end{figure*}
Using the high bandwidth and spatial resolution of the device we are able to image intricate structures on printed circuit boards (PCBs).
Figure\,\ref{fig:fig5} shows a miniature coat-of-arms of the city of Mainz as imaged by the eddy-current based method. The coat-of-arms is printed on a PCB, made out of copper with a 35\,$\mu$m thickness.
%\begin{figure*}[t]
%\centering
%\includegraphics[width=\textwidth]{dotsbw.png}
%\caption{(a) Fluorescence as a function of magnetic field at different alignments between the NV and magnetic field axis.
%(b) Lock-in amplifier R component for the same alignments %between the NV and the magnetic field axis .}
%\label{fig:fig5}
%\end{figure*}

%Figure\,\ref{fig:fig5}\ shows the imaging results of a sample with multiple 1\,mm diameter, 35\,$\mu m$ thick aluminum dots. Both the dots and the distance between them, which is less than 1\,mm, is clearly visible.  

\section*{Sensitivity}
To calculate the sensitivity of the device we have to make assumptions about the sample to be imaged. 
We assume a primary field of $B_{prim.}$\,=\,91\,$\mu$T with frequency of 3.5\,MHz, a cylindrical material with radius $r_0$\,=\,1\,mm and height $h$\,=\,0.1\,mm (volume $\sim$\,0.3\,mm$^3$). The sensor is placed at a distance $d$\,=\,0.5\,mm from the material. We calculate the minimum conductivity $\sigma$ that the material should have in order to produce a field strong enough to be detected with our magnetic field sensor (current sensitivity $\sim$ 10\,$\mu$T/$\sqrt[]{Hz}$).

The secondary magnetic field can be calculated using the expression for the electromotive force (EMF)
\begin{equation}
U\,=\,-\frac{\Delta \Phi}{\Delta t}=-2 E\pi r,
\end{equation}
where $\Delta \Phi$ is the magnetic flux and 1/$\Delta t$\,=\,$\omega$\,=\,2$\pi$f, f being the modulation frequency of $B_{prim.}$, $E$ is the electric field and $r$ is the distance from the center of the imaged sample. Using Biot-Savart's law, the field produced by the eddy-currents (i.e. secondary field, $B_{sec.}$)
and the minimum $\sigma$ able to be detected can be calculated.
%is:
%\begin{equation}
%B_{sec.}(r)=\frac{\mu_0}{4\pi}\int\frac{I\,dI\times r'}{|r'|^3},
%\end{equation}
%where $\mu_0$\,=\,4$\pi\times 10 ^{-7}$\,H/m is the vacuum permeability, $I$\,=\,$\sigma$\,dr\,dh and $r'$=$\sqrt[]{r^2+(d+h)^2}$.
%Substituting for the geometry we assumed before, we calculate from Eq.\,(2) that 
For the example presented it would be $\sim$\,8$\times 10 ^{5}$\,S/m/$\sqrt{Hz}$. Enough to detect small pieces of materials like copper and aluminum, ($\sigma_{copper}$\,=\,5.96$\times$10$^7$\,S/m, $\sigma_{aluminium}$\,=\,3.77$\times$10$^7$\,S/m) will be visible with our sensor.

\section*{Discussion}

We have demonstrated eddy-current imaging using NV
centers in diamond. The magnetic sensitivity is achieved with an all-optical method that is robust to misalignment and makes use of the NV-NV cross-relaxation feature in the PL as function of the magnetic field.
The bandwidth extends to 3.5\,MHz. The spatial resolution is 348\,$\mu$m\,$\pm$\,2\,$\mu$m. The sensitivity of the device is calculated as 8$\times 10 ^{5}$\,S/m\,$\sqrt[]{\textrm{Hz}}$ at 3.5\,MHz, for a cylindrical sample with radius $r_0$\,=\,1\,mm and height $h$\,=\,0.1\,mm (volume $\sim$\,0.3\,mm$^3$) at a distance of 0.5\,mm.
Combining the above characteristics our device can be used, to distinguish between different materials, %(copper and aluminum\cite{Supplemental}),
detect structural defects, 
%(9\,mm diameter ring with 1\,mm slit\cite{Supplemental}),
and image complex structures on PCBs.
%\cite{Supplemental}.
The technique can be further improved by closer proximity to the imaged sample, which can be achieved using a thinner diamond or one with a shallow implanted NV layer. 
Compared to measurements with atomic vapor cells, the current NV-based method has superior bandwidth and spatial resolution. 
Comparing with coil-based commercial devices, it features superior bandwidth (e.g. Zetec, MIZ-22, bandwidth 50\,Hz to 2\,MHz) and sensitivity (5$\times 10 ^{5}$\,S/m).

With its high bandwidth and spatial resolution, if combined with state-of-the-art NV sensors\cite{Chatzidrosos,Barry2016} the sensitivity (on the order of 10 pT/$\sqrt[]{Hz}$) would be sufficient
%(see supplementary\cite{Supplemental})
for applications in biomedicine to distinguish between different tissues, or even healthy and unhealthy tissues, because of their different conductivities (ranging from 0.5\,mS/cm to 13.7\,mS/cm)\cite{doi:10.1080/16070658.1981.11689230} . The device's robustness to misalignment and small size would allow it to be implemented on a hand-held, small and portable endoscopic sensor.

\begin{acknowledgements}

The authors acknowledge support by the EU FET-OPEN Flagship Project ``ASTERIQS'' (action \#\,820394), 
the German Federal Ministry of Education and Research (BMBF) within the Quantumtechnologien program (FKZ 13N14439),
and the DFG through the DIP program (FO 703/2-1). 
GC acknowledges support by the internal funding of JGU. 
LB is supported by a Marie Curie Individual Fellowship within the second Horizon 2020 Work Programme.
HZ is a recipient of a fellowship through GRK
Symmetry Breaking (DFG/GRK 1581). We thank N. L. Figueroa for fruitful
discussions.
\end{acknowledgements}
\section*{References}

\bibliographystyle{ieeetr}

\bibliography{literature}

\begin{thebibliography}{10}

\bibitem{Marmugi2016}
L.~Marmugi and F.~Renzoni, ``Optical magnetic induction tomography of the
  heart,'' {\em Scientific Reports}, vol.~6, pp.~10--12, Apr 2016.

\bibitem{Deans2018}
C.~Deans, L.~Marmugi, and F.~Renzoni, ``Active underwater detection with an
  array of atomic magnetometers,'' {\em Applied Optics}, vol.~57,
  pp.~2346--2351, 2018.

\bibitem{Hussain2015}
S.~Hussain, L.~Marmugi, C.~Deans, and F.~Renzoni, ``Electromagnetic imaging
  with atomic magnetometers: a novel approach to security and surveillance,''
  {\em SPIE 9823, Detection and Sensing of Mines, Explosive Objects, and
  Obscured Targets XXI, 98230Q}, vol.~9823, May 2016.

\bibitem{Angelo2015}
G.~D'Angelo and S.~Rampone, ``Shape-based defect classification for non
  destructive testing,'' pp.~406--410, 08 2015.

\bibitem{Arne2016}
A.~Wickenbrock, N.~Leefer, J.~W. Blanchard, and D.~Budker, ``Eddy current
  imaging with an atomic radio-frequency magnetometer,'' {\em Applied Physics
  Letters}, vol.~108, p.~183507, 2016.

\bibitem{Balasubramanian2008}
G.~Balasubramanian, I.~Y. Chan, R.~Kolesov, M.~Al-Hmoud, J.~Tisler, C.~Shin,
  C.~Kim, A.Wojcik, P.~R. Hemmer, A.~Krueger, T.~Hanke, A.~Leitenstorfer,
  R.~Bratschitsch, F.~Jelezko, and J.~Wrachtrup, ``Nanoscale imaging
  magnetometry with diamond spins under ambient conditions,'' {\em Nature},
  vol.~455, no.~7213, pp.~648--651, 2008.

\bibitem{Maze2008}
J.~R. Maze, P.~L. Stanwix, J.~S. Hodges, S.~Hong, J.~M. Taylor, P.~Cappellaro,
  L.~Jiang, M.~V.~G. Dutt, E.~Togan, A.~S. Zibrov, A.~Yacoby, R.~L. Walsworth,
  and M.~D. Lukin, ``Nanoscale magnetic sensing with an individual electronic
  spin in diamond,'' {\em Nature}, vol.~455, no.~7213, pp.~644--647, 2008.

\bibitem{Rittweger2009}
E.~Rittweger, K.~Y. Han, S.~E. Irvine, C.~Eggeling, and S.~W. Hell, ``Sted
  microscopy reveals crystal colour centres with nanometric resolution,'' {\em
  Nat Photon}, vol.~3, no.~3, pp.~144--147, 2009.

\bibitem{Barry2016}
J.~F. Barry, M.~J. Turner, J.~M. Schloss, D.~R. Glenn, Y.~Song, M.~D. Lukin,
  H.~Park, and R.~L. Walsworth, ``Optical magnetic detection of single-neuron
  action potentials using quantum defects in diamond,'' {\em PNAS}, vol.~113,
  p.~14133, 2016.

\bibitem{Chatzidrosos}
G.~Chatzidrosos, A.~Wickenbrock, L.~Bougas, N.~Leefer, T.~Wu, K.~Jensen,
  Y.~Dumeige, and D.~Budker, ``Miniature cavity-enhanced diamond
  magnetometer,'' {\em Phys. Rev. Applied}, vol.~8, p.~044019, 2017.

\bibitem{Dumeige2013}
Y.~Dumeige, M.~Chipaux, V.~Jacques, F.~Treussart, J.-F. Roch, T.~Debuisschert,
  V.~M. Acosta, A.~Jarmola, K.~Jensen, P.~Kehayias, and D.~Budker,
  ``Magnetometry with nitrogen-vacancy ensembles in diamond based on infrared
  absorption in a doubly resonant optical cavity,'' {\em Phys. Rev. B},
  vol.~87, p.~155202, Apr 2013.

\bibitem{Armstrong1}
S.~Armstrong, L.~J. Rogers, R.~L. McMurtrie, and N.~B. Manson, ``{NV-NV}
  electron–electron spin and {NV-NS} electron — electron and
  electron-nuclear spin interaction in diamond,'' {\em Physics Procedia},
  vol.~3, no.~4, pp.~1569 -- 1575, 2010.
\newblock Proceedings of the Tenth International Meeting on Hole Burning,
  Single Molecule and Related Spectroscopies: Science and Applications-HBSM
  2009.

\bibitem{Hall2016}
L.~T. Hall, P.~Kehayias, D.~A. Simpson, A.~Jarmola, A.~Stacey, D.~Budker, and
  L.~C.~L. Hollenberg, ``Detection of nanoscale electron spin resonance spectra
  demonstrated using nitrogen-vacancy centre probes in diamond,'' {\em Nat
  Commun}, vol.~7, 2016.

\bibitem{wickenbrock2016microwave}
A.~Wickenbrock, H.~Zheng, L.~Bougas, N.~Leefer, S.~Afach, A.~Jarmola, V.~M.
  Acosta, and D.~Budker, ``Microwave-free magnetometry with nitrogen-vacancy
  centers in diamond,'' {\em Applied Physics Letters}, vol.~109, 2016.

\bibitem{Ivanov2016}
S.~V. Anishchik and K.~L. Ivanov, ``{Level-crossing spectroscopy of
  nitrogen-vacancy centers in diamond: sensitive detection of paramagnetic
  defect centers},'' {\em Phys. Rev. B}, vol.~96, p.~115142, 2017.

\bibitem{Wood2016_2}
J.~D.~A. Wood, D.~A. Broadway, L.~T. Hall, A.~Stacey, D.~A. Simpson, J.-P.
  Tetienne, and L.~C.~L. Hollenberg, ``Wide-band nanoscale magnetic resonance
  spectroscopy using quantum relaxation of a single spin in diamond,'' {\em
  Phys. Rev. B}, vol.~94, p.~155402, 2016.

\bibitem{Zheng2017}
H.~Zheng, G.~Chatzidrosos, A.~Wickenbrock, L.~Bougas, R.~Lazda, A.~Berzins,
  F.~H. Gahbauer, M.~Auzinsh, R.~Ferber, and D.~Budker, ``Level anti-crossing
  magnetometry with color centers in diamond,'' {\em Proc. SPIE}, vol.~10119,
  2017.

\bibitem{Marmugi22018}
L.~Marmugi, C.~Deans, and R.~Renzoni, ``Atomic magnetometry-based
  electromagnetic imaging of low-conductivity semiconductors,'' {\em
  arXiv:1805.05743}, 2018.

\bibitem{doi:10.1080/16070658.1981.11689230}
K.~Foster and J.~Schepps, ``Dielectric properties of tumor and normal tissues
  at radio through microwave frequencies,'' {\em Journal of Microwave Power},
  vol.~16, no.~2, pp.~107--119, 1981.

\end{thebibliography}

\end{document}